\documentclass[12pt,a4paper]{article}
\usepackage[centertags]{amsmath}
\usepackage{amssymb}
\usepackage{epsfig} 
\usepackage{amsmath} 
\usepackage{graphicx} 
\usepackage{jheppub}

\newcommand{\beq}{\begin{equation}}
\newcommand{\eeq}{\end{equation}}
\newcommand{\bea}{\begin{eqnarray}}
\newcommand{\eea}{\end{eqnarray}}
\newcommand{\ba}{\begin{array}}
\newcommand{\ea}{\end{array}}
\newcommand{\bi}{\begin{itemize}}
\newcommand{\ei}{\end{itemize}}
\newcommand{\bn}{\begin{enumerate}}
\newcommand{\en}{\end{enumerate}}
\newcommand{\bc}{\begin{center}}
\newcommand{\ec}{\end{center}}
\renewcommand{\l}{\left}
\renewcommand{\r}{\right}

\newcommand{\ol}{\overline}

\newcommand{\Ga}{\Gamma}
\newcommand{\De}{\Delta}
\newcommand{\Om}{\Omega}

\newcommand{\al}{\alpha}

\newcommand{\ga}{\gamma}

\newcommand{\la}{\lambda}

\newcommand{\si}{\sigma}

\newcommand{\nl}{\nonumber\\}

\newcommand{\eq}[1]{Eq.~(\ref{#1})}

\newcommand{\GeV}{\mathinner{\mathrm{GeV}}}

\title
{\Large \bf
Search for the Higgs portal to a singlet fermionic dark matter at the LHC}

\author{Seungwon Baek, P. Ko, Wan-Il Park\\
{\it School of Physics, KIAS, \\
85 Hoegiro Dongdaemun-gu, Seoul 130-722, Korea}
}

\emailAdd{sbaek1560@gmail.com}
\emailAdd{pko@kias.re.kr}
\emailAdd{wipark@kias.re.kr}

\abstract{We consider a simple extension of the standard model with a singlet 
fermionic dark matter. Its thermal relic density can be easily 
accommodated by a real singlet scalar messenger that mixes with 
the standard model Higgs boson.  The model can change significantly 
the Higgs signals at the LHC via sizable invisible decays of two 
Higgs-like scalar bosons.
After imposing the constraints from the electroweak precision tests,
colliders and dark matter search experiments, 
one concludes that two or one or none of the two
Higgs bosons, depending on the mass relations among
two scalar bosons and the dark matter fermion and their couplings.
In particular, if a standard model Higgs-like scalar boson is discovered 
around 120--125 GeV region at the LHC, it would be almost impossible to 
find the second Higgs-like boson since it is mostly a singlet scalar, 
whether it is heavier or lighter.
This model can be further tested by direct dark matter search experiments.
}

\begin{document}
\maketitle

\section{Introduction}

The Standard Model (SM) of particle physics has been very successful for describing 
many phenomena observed at various experiments, except for neutrino oscillation,
nonbaryonic cold dark matter (DM) and  baryon number asymmetry of the universe, 
as well as some theoretical shortcomings such as  hierarchy problem, strong CP 
problem, {\it etc}. 
At present, the only part of the SM which experiments could not unveil is 
the Higgs sector, namely electroweak symmetry breaking (EWSB) sector.
The absence of signal at LEP experiment puts a lower bound on the mass of the 
SM Higgs particle at about $114.4 \GeV$~\cite{Barate:2003sz}.
However, the most recent results from ATLAS and CMS at the LHC have 
excluded  the SM Higgs boson in the mass range of $141 - 476$ GeV at 
$95\%$ CL ~\cite{ATLAS}.  Hence, the only remaining window for the light Higgs mass 
is $114.4 - 141 \GeV$.  This is consistent with the electroweak precision 
test (EWPT) which strongly favors the light SM Higgs boson. 

When we consider Higgs physics, it is very important to notice that 
dark matter can easily modify Higgs phenomenology. 
For example, in the real singlet scalar DM model with $Z_2$ symmetry, 
this happens through the invisible Higgs decay into a pair of DM's: 
$H \rightarrow D D$ where $D$ is a real singlet scalar dark matter.  
In this model lagrangian, the stability of the real singlet scalar $D$ is 
guaranteed by an ad hoc $Z_2$ symmetry $D \rightarrow -D$.  
There are a number of works in this direction \cite{Silveira:1985rk,
hep-ph/0702143,hep-ph/0011335} (see Refs.~
\cite{Kanemura:2010sh,Mambrini:2011ik,Raidal:2011xk,He:2011de,Ma:2011kc,Gupta:2011gd} 
for recent discussions). 
One can consider a complex scalar cold dark matter (CDM), for which one has qualitatively 
similar results~\cite{arXiv:0811.0393}.

One can also consider singlet dark matter with higher spins: 
singlet fermion or singlet vector boson dark matter.  
There are some works in this direction, sometimes under the name of Higgs-portal
dark matter \cite{KYLee:2008,Kanemura:2010sh,Kanemura:2011vm,Lindner:2011it,
arXiv:1110.4405,Kanemura:2011mw,Lebedev:2011iq}, many of which work in the effective 
lagrangian approaches with the SM particles and the singlet dark matters.   
For the case of a singlet fermionic dark matter, Ref.~\cite{KYLee:2008} 
employed a renormalizable lagrangian similar to our present work.  
But overall phenomenology of the scalar sector such as the EWPT was not discussed.  
In this paper, we will make more emphasis on the EWPT, the Higgs phenomenology 
at the LHC and the interplay between the DM sector and two Higgs-like
scalar bosons, which makes the first attempt to consider all the relevant 
phenomenological aspects related with the singlet fermion dark matter and 
the Higgs bosons.

If we consider a singlet fermion CDM scenario at a renormalizable 
lagrangian level, there appears an additional singlet scalar  $s$ which 
plays a role of messenger between the SM sector and the DM $\psi$.  
In this scenario, there will be at least two scalars $H_1$ and $H_2$,  
the mixtures of the SM Higgs boson  $h$ and the singlet scalar $s$
(see Refs.~\cite{Profumo:2007,Barger:2007}
for recent discussions on the singlet scalar extension of the SM). 
Since the scalar boson spectrum is qualitatively different from the singlet 
scalar DM scenario where there is only one SM Higgs boson, 
it is important to analyze the singlet fermion CDM scenario 
in a more quantitative way, and understand the generic signatures
at the LHC and at other DM search experiments.  It is also important to 
notice that the effective lagrangian approach with the SM Higgs boson can 
miss some important features of dark matter models such as our model with 
two Higgs-like scalar bosons.  For example, we will observe that there is a generic
cancellation of the Higgs boson contributions in the direct detection cross section
between the DM and a nucleon, which can not be seen in the effective lagrangian
approach.
 
In the singlet fermion DM model with a real singlet scalar messenger,  
the Higgs phenomenology can be modified in some different reasons:
\begin{itemize}
\item Mixing between $h$ and $s$ makes the physical Higgs bosons $H_1$
and $H_2$ have reduced couplings with the SM fermions and the SM weak 
gauge bosons.
\item $H_{i=1,2}$ can decay into a pair of CDM's 
$H_i \rightarrow \psi \overline{\psi}$  if kinematically allowed.
\item $H_2 \rightarrow H_1 H_1$ can contribute to the decay width of the 
heavier scalar  boson $H_2$, if kinematically allowed.
\item The first two dilution factors make the production and the detection of 
two Higgs bosons more difficult than the SM Higgs boson case.  There is an  
ample parameter space where one of the scalar bosons or both can not be
discovered at the LHC even at 10 fb$^{-1}$ (see Fig.~10). 
\item Presence of two scalars $H_1$ and $H_2$ relaxes the strong constraint
from the direct detection cross section, whereas the couplings of $H_i - \psi -
\overline{\psi}$ can be still large and the 
$B ( H_i \rightarrow \psi \overline{\psi} )$ can be substantial.
This makes an important difference from  the real singlet scalar CDM with 
$Z_2$,  in which the direct detection cross section puts a strong constraint 
on the  $B ( H \rightarrow D D)$.
\item Even if no SM Higgs boson is found at the LHC in the end, 
it does not necessarily imply that the perturbative unitarity of $V_L V_L$ 
scattering amplitude is broken, or there should be new strongly interacting 
EWSB sector.  Our scenario with $m_1 \sim m_2$ and $r_1 , r_2 \sim 0$
can describe such a situation, and would be perfectly fine. 
\end{itemize}

These features are also qualitatively true in other hidden sector DM models, 
regardless of  strongly or weakly interacting hidden sector 
\cite{Hur:2007uz,arXiv:1103.2571,ko_talks,Englert:2011yb}. 
In case the hidden sector gauge interaction is a confining gauge theory
like QCD, there will appear Nambu-Goldstone (NG) bosons at low energy, 
which could be the CDM. And new composite scalar bosons (similar to 
$\sigma$ meson in QCD) in the  hidden sector can mix with the SM Higgs 
boson or a real singlet messenger scalar $S$. 
The number of neutral scalar bosons will depend on the number of hidden sector
quark flavors, and becomes model dependent.  One crucial difference of this type 
of scenario from the multi-Higgs doublet models is that there will be only
neutral scalar bosons, and not any charged scalar bosons.   

Finally, in a class of models where the dark matter sector and the SM 
sector carry a new gauge symmetry in common, one could have similar 
phenomena.  If the new gauge boson gets its mass from
spontaneous symmetry breaking due to a new scalar field $\phi$, 
there will be a generic mixing between this new scalar field and the SM 
Higgs field, via $\phi^\dagger \phi H^\dagger H$ interaction term 
and nonzero VEV's of $H$ and $\phi$.
For example, two of the present authors studied the gauged $U(1)_{L_\mu 
- L_\tau}$ extension of the SM  and explained the PAMELA excess, without 
any conflict with many constraints from the low energy experiments, colliders
and astrophysical observations \cite{baek_ko}. 

All these models have a similar consequences for the Higgs phenomenology, 
namely more than one Higgs-like neutral scalar bosons with substantial 
invisible branching ratios, and improves the fit to the electroweak precision data. 
It would not be easy to distinguish one model from another using the 
experimental data, and it is beyond the scope of this paper to attempt 
such a study here. 

In this paper, we study the singlet fermion dark matter model to see if it can 
explain the recent LHC data while fulfilling other observational and cosmological 
requirements,  and if Higgs particle(s) can be discovered from the future 
data-accumulation at the LHC,  keeping in mind that there could be other models
which might have similar observational consequences. Our model is one of the
simplest extension of the SM with a singlet fermion dark matter and thus 
could serve as  a good starting point for phenomenological analysis and 
the analysis strategy can be applied to  other models for more general study. 
  
This paper is organized as follows.  
In Sec.~2, we define the singlet fermion CDM model with two Higgs-like 
scalar bosons.  In Sec.~3, we calculate the decay rates for 
$H_i \rightarrow \psi \overline{\psi}$, 
and discuss the relevant constraints from colliders, electroweak precision 
tests and dark matter phenomenology.  In Sec.~4, the phenomenology of 
two Higgs scalar bosons is  discussed for three different cases:  
$m_1 (\sim 120 \GeV) \ll m_2$, 
$m_1 \sim m_2 \sim 120$ GeV,  and $m_1 \ll m_2 (\sim 120 \GeV)$, 
assuming the mass of one Higgs boson is 120 GeV, and 
discuss the detectability of two Higgs boson at the LHC.  
Finally the results and their implications are summarized in Sec.~5.

\section{The model}

We consider an extension of the SM, adding a singlet Dirac
dark matter\footnote{We get similar results for a Majorana dark matter
case.}   $\psi$ and a singlet scalar $S$. 
The singlet fermion DM $\psi$ is assumed to live in the hidden sector, and 
communicates with the SM sector via the scalar $S$. 
Then, the model lagrangian has 3 pieces, 
the hidden sector and Higgs portal terms in addition to the SM lagrangian: 
\bea
 {\cal L} = {\cal L}_{\rm SM} + {\cal L}_{\rm hidden} + {\cal L}_{\rm portal},
\label{eq:Lag}
\eea
where 
\bea
{\cal L}_{\rm hidden} &=& {\cal L}_S + {\cal L}_\psi - \la S \ol{\psi} \psi, \nl
{\cal L}_{\rm portal} &=& - \mu_{HS} S H^\dag H -{\la_{HS} \over 2} S^2 H^\dag H,
\label{eq:Lag2}
\eea
with 
\bea
{\cal L}_S &=&
{1 \over 2} (\partial_\mu S \partial^\mu S - m_S^2 S^2) 
-\mu_S^3 S - {\mu_S^\prime \over 3} S^3  - {\la_S \over 4} S^4, \nl
{\cal L}_\psi &=&
\ol{\psi} ( i \not \partial - m_{\psi_0} ) \psi.
\label{eq:Lag3}
\eea
The model without the singlet fermion DM, namely the SM plus an additional 
singlet scalar field $S$ has been studied in detail in ~\cite{Profumo:2007,Barger:2007}.
Note that a real scalar singlet dark matter model can be obtained 
if we remove the fermionic DM $\psi$ and impose $Z_2$ symmetry: 
$S\rightarrow -S$.

The Higgs potential has three parts: the SM, the hidden sector and
the portal parts
\bea
 V_{\rm Higgs} = V_{\rm SM} + V_{\rm hidden} + V_{\rm portal},
\eea
where $V_{\rm hidden}, V_{\rm portal}$ can be read from (\ref{eq:Lag2}), (\ref{eq:Lag3}) and
\bea
 V_{\rm SM} = -\mu_H^2 H^\dag H + \la_H (H^\dag H)^2.
\eea
In general the Higgs potential develops nontrivial vacuum expectation values,
$v_H$ and $v_S$, and we can expand $H$ and $S$ as
\bea
  H =  \l(
 \begin{array}{c} G^+ \\ {1 \over \sqrt{2} }  (v_H + h + i G^0)  \end{array} \r), 
 \quad
 S = v_S + s,
\eea
where $G^+$ and $G^0$ are the Goldstone bosons and $h$ and $s$ are physical 
scalar fields after $H$ and $S$ develops nonzero VEV's.
Assuming all the Higgs sector parameters are real, that is, there is no CP violation in the Higgs sector, we obtain
\bea
 \mu_H^2 &=& \la_H v_H^2 + \mu_{HS} v_S + {1 \over 2} \la_{HS} v_S^2, \nl
 m_S^2 &=& -\frac{\mu_S^3}{v_S} - \mu_S^\prime v_S - \la_S v_S^2 -\frac{\mu_{HS} v_H^2}{2 v_S}
 -{1 \over 2} \la_{HS} v_H^2,
\eea
from the tadpole conditions. The quartic couplings can be traded for the
Higgs mass parameters
\bea
 \la_H &=& \frac{m_{hh}^2}{2 v_H^2}, \nl
 \la_{HS} &=& \frac{m_{hs}^2 -\mu_{HS} v_H}{v_S v_H}, \nl
 \la_{SS} &=& \frac{m_{ss}^2 + \mu_S^3/v_S - \mu_S^\prime v_S -\mu_{HS} v_H^2/(2 v_S)}{2 v_S^2}.
\eea
Now the Higgs mass matrix can be diagonalized by introducing mixing angle $\al$ so that
\bea
 M^2_{\rm Higgs} \equiv
 \l(\begin{array}{cc} m_{hh}^2 & m_{hs}^2 \\ m_{hs}^2 & m_{ss}^2 \end{array}\r)
 \equiv \l(\begin{array}{cc} \cos\al & \sin\al \\ -\sin\al & \cos\al \end{array}\r)
\l(\begin{array}{cc} m_1^2 & 0 \\ 0 & m_2^2 \end{array}\r)
\l(\begin{array}{cc} \cos\al & -\sin\al \\ \sin\al & \cos\al \end{array}\r).
\eea
The mass eigenstates with masses $m_1$ and $m_2$ are expressed in terms of the SM Higgs $h$ and the singlet $s$
as
\bea
 H_1 &=& h \cos\al - s \sin\al, \nl
 H_2 &=& h \sin\al + s \cos\al.
\label{eq:mass_weak}
\eea
We take
\bea
 m_1, \quad m_2, \quad \al, \quad v_S, \quad \mu_S, \quad 
\mu_S^\prime,\quad  \mu_{HS}
\eea
as free parameters for the Higgs sector.
We have two additional free parameters, the DM mass and its 
coupling to the singlet scalar $S$:
\bea
 m_{\psi} (\equiv m_{\psi_0} + \la v_S), \quad \la.
\eea
Therefore, we have introduced 9 more new parameters in total, compared 
with the SM lagrangian.
These new parameters are constrained by various theoretical, experimental 
and observational data: perturbative unitarity of gauge boson scattering
amplitudes, EWPT, collider searches for Higgs boson(s), 
DM relic density, DM direct detection experiments, {\it etc}.

\section{Constraints}

In this section we consider the following constraints on the model parameters: 
\begin{itemize}
\item the perturbative unitarity condition on the Higgs 
sector~\cite{bwlee,Englert:2011yb},
\item the LEP bound on the SM Higgs boson mass~\cite{Barate:2003sz},
\item the oblique parameters $S$, $T$ and $U$ obtained from the EWPT
~\cite{Peskin:1990zt,Maksymyk:1993zm},
\item the observed CDM density, 
$\Omega_{\rm CDM} h^2 =0.1123 \pm 0.0035$~\cite{Jarosik:2010iu}, 
which we assume is saturated by the thermal relic $\psi$,
\item the upper bound on the DM-proton scattering cross section 
obtained by the XENON100 experiment~\cite{Aprile:2011hi}.
\end{itemize}
Note that the first three constraints are independent of the dark matter sector, 
and they apply to the SM plus a singlet scalar model without dark matter as well.

\subsection{Perturbative unitarity of gauge boson scattering amplitudes}

The perturbative unitarity of  scattering amplitudes for longitudinal weak 
gauge bosons in our model  requires~\cite{bwlee,Englert:2011yb}, 
\beq
\langle m^2 \rangle \equiv m_1^2 \cos^2 \alpha  + m_2^2 \sin^2 \alpha  \leq 
\frac{4 \pi \sqrt{2}}{3 G_F} \approx \left( 700 \GeV \right)^2
\label{eq:unitarity}
\eeq 
This is a rather weak constraint compared with other constraints that 
will be described subsequently, and thus does not play an important role.

\subsection{LEP bound}
The LEP data severely constrains the ratio of Higgs-$Z$-$Z$ 
coupling strength to that of the SM, 
$\xi^2 \equiv (g_{HZZ}/g_{HZZ}^{\rm SM})^2$~\cite{Barate:2003sz}.
For example, as shown in Fig.~10(a) in Ref.~\cite{Barate:2003sz}, 
if $m_H$ is less than 90 GeV, the $\xi^2$ should be less than 
$\mathcal{O}(0.1)$.

The signal strength or ``the reduction factor'' in the event number 
of a specific final state SM particles,  $X_{\rm SM}$, in the Higgs 
boson decays is defined as
\beq
r_i \equiv \frac{\si_{H_i} B_{H_i \to X_{\rm SM}}}{\si^{\rm SM}_{H_i} 
B^{\rm SM}_{H_i \to X_{\rm SM}}}\ \ (i=1,2),
\eeq
where $\si_{H_i}$ and $B_{H_i \to X_{\rm SM}}$ are the production cross 
section of $H_i$, and the branching ratio of $H_i \to X_{\rm SM}$ respectively,  
while $\si^{\rm SM}_{H_i}$ and $B^{\rm SM}_{H_i \to X_{\rm SM}}$ are 
the corresponding quantities of the SM Higgs with mass $m_i$.
Then we find
\bea
 r_1 &=&  \frac{c_\al^4 \Ga_{H_1}^{\rm SM}}{c_\al^2 \Ga_{H_1}^{\rm SM}+s_\al^2 \Ga_{H_1}^{\rm hid}}, \nl
 r_2 &=&  \frac{s_\al^4 \Ga_{H_2}^{\rm SM}}{s_\al^2 \Ga_{H_2}^{\rm SM}+c_\al^2 \Ga_{H_2}^{\rm hid}+\Ga_{H_2\to H_1 H_1}},
\label{eq:rf}
\eea
where $c_\al \equiv \cos \al,\, s_\al \equiv \sin\al$.
The $\Ga_{H_i}^{\rm SM}$ denotes the total decay width of the SM Higgs boson 
with mass $m_i$ and the $\Ga_{H_i}^{\rm hid}$ is that of $H_i \to \psi \overline{\psi}$
(the invisible decay modes of $H_i$'s).
Note that the signal strength $r_i$ is reduced by $c_\al (s_\al)$ in the 
production cross section due to the mixing between $h$ and $s$, 
as shown in (\ref{eq:rf}), even if the invisible mode 
($H_i \rightarrow \psi \overline{\psi}$) or the Higgs-splitting  mode 
($H_2 \to H_1 H_1$) is kinematically forbidden in the Higgs decay.
In other words, a reduced signal of the Higgs boson at the LHC would 
be a generic signature of the mixing of the SM Higgs boson with 
extra singlet scalar boson(s).

For the numerical analysis, we take the SM-like Higgs mass to be 120~GeV
as a benchmark value, for which the SM Higgs decays dominantly into 
$b \bar{b}$, and we obtain
\beq
\Gamma_{h \to \rm SM} \simeq 0.04 \GeV.
\label{eq:GamH_SM}
\eeq
This can be compared with the hidden sector decay width
\beq
\Ga_{H_i}^{\rm hid}
= \frac{\lambda^2 m_i}{8 \pi}  \left( 1 - \frac{4 m_\psi^2}{m_i^2} \right)^{3/2}.
\eeq
For $\lambda = 1$, $m_i = 120$ GeV and $m_\psi = 55$ GeV, we get
$\Ga_{H_i}^{\rm hid} = 0.3$ GeV which is much larger than (\ref{eq:GamH_SM}).
This may impose serious problems in searching for Higgs at the LHC as
will be discussed below.

For $m_2 > 2 m_1$, the Higgs splitting mode $H_2 \rightarrow H_1 H_1$ 
will open, which would generate very peculiar signals at colliders such as 
$H_2 \to H_1 H_1 \to b b \bar{b} \bar{b}, b \bar{b} 
\tau^+ \tau^-,  \tau^+ \tau^+ \tau^- \tau^-$, 
hence could be a target at future collider experiments.

\subsection{The oblique parameters: $S, T, U$}

The extended Higgs sector gives extra contribution to the gauge boson 
self-energy diagrams, as the SM Higgs boson does.
This can affect the EWPT leading to  the constraints on the oblique parameters, 
$S, T$ and $U$, by the Higgs sector
\footnote{If the Higgs masses are not much larger than electroweak scale, 
$V$ and $W$ parameters should be also taken into account.
However, their contributions are smaller than those of $S$, $T$ and $U$ 
parameters and makes no much difference.
Hence we ignore the effects on $V$ and $W$ in our argument.}.
Since the newly added singlet scalar $S$ is electrically neutral, 
$\Pi_{\ga\ga}, \Pi_{\ga Z}$ do not change from the SM predictions, 
and we only need to calculate $W$ and $Z$ boson self-energy diagrams, 
$\Pi_{WW}, \Pi_{ZZ}$.

It is straight-forward to get $\De X \equiv X - X^{SM}, (X=S, T, U)$,
\bea
\De T &=& \frac{3}{16 \pi s_Z^2} \Bigg[
\cos^2\al \l\{f_T\l(m_1^2 \over M_W^2\r)-{1 \over c_Z^2}f_T\l(m_1^2 \over M_Z^2\r)\r\} \nl
&+&\sin^2\al \l\{f_T\l(m_2^2 \over M_W^2\r)-{1 \over c_Z^2}f_T\l(m_2^2 \over M_Z^2\r)\r\}
- \l\{f_T\l(m_h^2 \over M_W^2\r)-{1 \over c_Z^2}f_T\l(m_h^2 \over M_Z^2\r)\r\} \Bigg], \nl
\De S &=& \frac{1}{2 \pi} \Bigg[ \cos^2 \al f_S\l(m_1^2 \over M_Z^2\r)
+ \sin^2 \al f_S\l(m_2^2 \over M_Z^2\r)
- f_S\l(m_h^2 \over M_Z^2\r) \Bigg], \nl
\De U &=& \frac{1}{2 \pi} \Bigg[ \cos^2 \al f_S\l(m_1^2 \over M_W^2\r)
+ \sin^2 \al f_S\l(m_2^2 \over M_W^2\r)
- f_S\l(m_h^2 \over M_W^2\r) \Bigg] - \De S .
\eea
The functions $f_T(x)$ and $f_S(x)$ are defined in the Appendix.
For the reference Higgs mass $m_h$ we fix $m_h = 120$ GeV.
The above expressions show that there is a symmetry in the $S, T$ and $U$
parameters in the simultaneous exchange of $ m_1, m_2$ and $\alpha$ such that  
\bea
 \De X(\al, m_1, m_2) = \De X({\pi \over 2} - \al, m_2, m_1).
\eea

\begin{figure}
\centering
\includegraphics[width=0.45\textwidth]{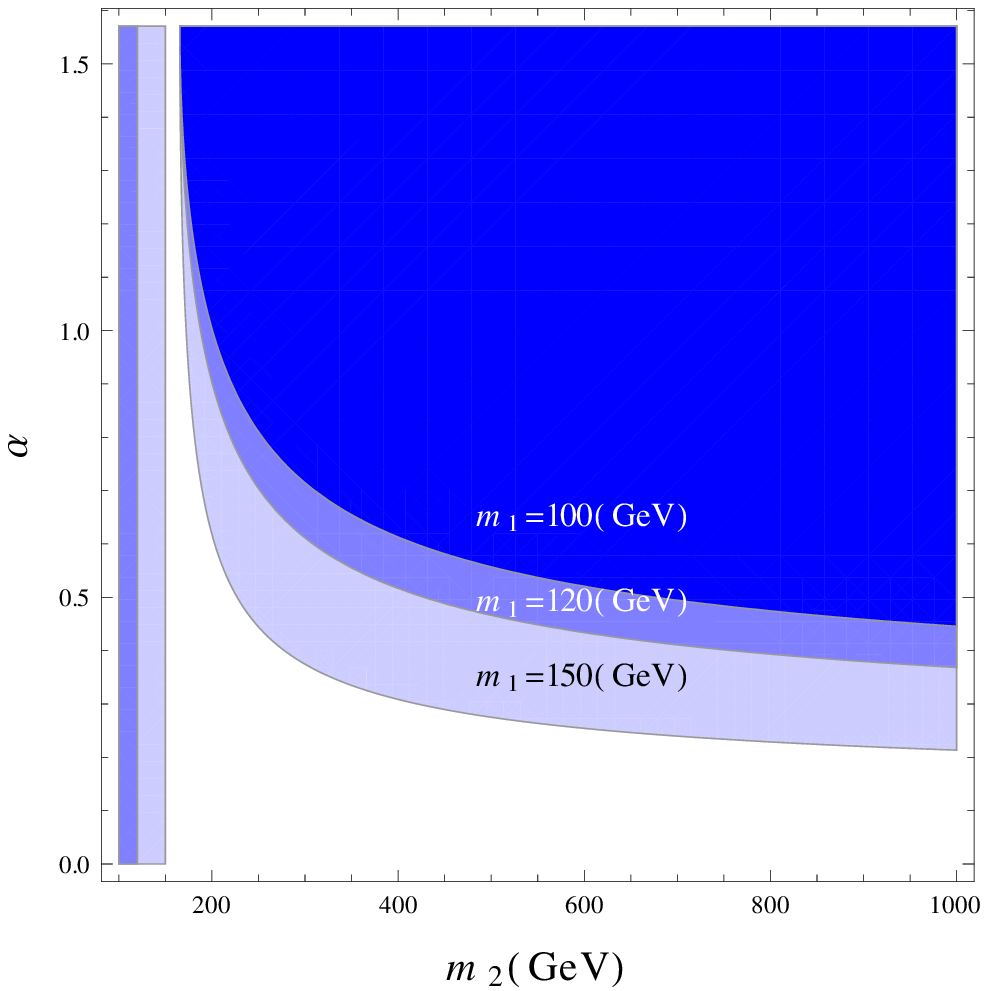}
\includegraphics[width=0.45\textwidth]{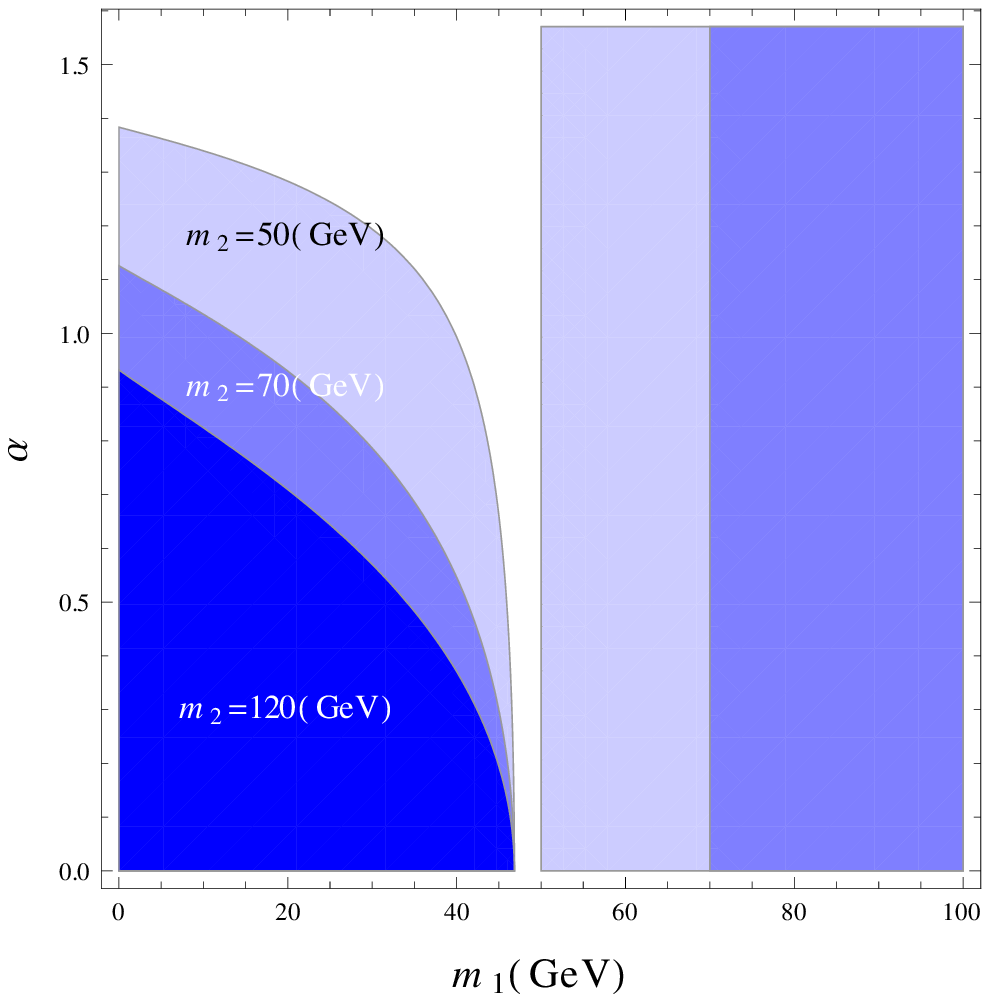}
\caption{
Left (Right) panel: Exclusion plot in $(m_{2(1)}, \al)$ plane 
with several choices of $m_{1(2)}$. 
The colored region for the given value of the Higgs mass is excluded by EWPT.
(Note that $m_2 > m_1$ by definition and only the region satisfying this 
relation is meaningful.)}
\label{fig:STU_m2_al}
\end{figure}
In order to obtain constraints on the parameters $m_1, m_2$ and $\al$ 
from $S, T, U$ parameters, in Fig.~\ref{fig:STU_m2_al} we show an exclusion 
plot in $(m_{2(1)}, \al)$ plane by the EWPT at 95\% CL~
\cite{Barger:2007,Baak:2011ze}.  The colored region for the given value 
$m_{1(2)}$ is  excluded in the left (right) panel. 
We can see that $H_2$ can be much heavier than ~$200 \GeV$ only when the $H_1$ 
is dominated by the SM component ({\it i.e.} $\al < \pi/4$) (see the left panel). 
The possibility  $m_1 < m_2 (\sim {\cal O}(100 \GeV)) $ is also allowed.
In this case $H_1$ can be lighter than $\sim 50 \GeV$ only when $H_2$ is SM-like
({\it i.e.} $\al > \pi/4$) (see the right panel.)

\begin{figure}
\centering
\includegraphics[width=0.7\textwidth]{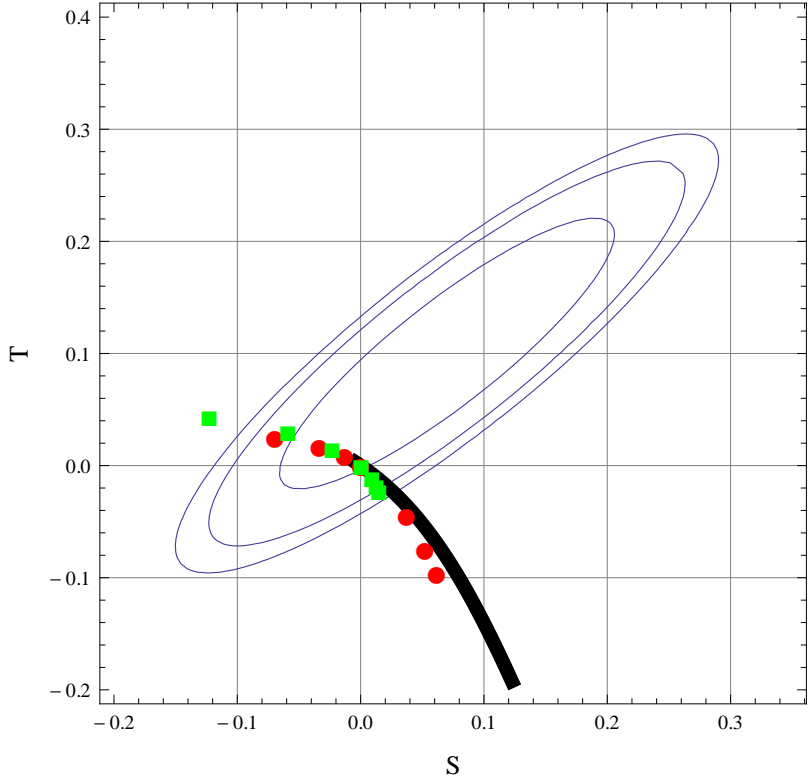}
\caption{
$(S,T)$ parameters in our model.   
The ellipses are (68, 90, 95)~\% CL contours.
The thick black curve shows the SM prediction with the Higgs boson mass 
in the region $(115,720)$ GeV.
The red, green dots correspond to $\alpha=45^\circ, 20^\circ$, respectively.
The dots are for $(m_1,m_2)(\GeV) =(30,120),(60,120),(90,120),(120,120),
(120,320),(120,520),(120,720)$ from above for each color.
}
\label{fig:STU}
\end{figure}

Fig.~\ref{fig:STU} shows that the EWPT constraint on our model is 
generically much less severe than on the SM. Since $\De U$ is very small, 
we assume $\De U=0$ for this plot.
The ellipses are (68, 90, 95)~\% CL contours~\cite{Baak:2011ze}.
The thick black curve shows the SM prediction with the Higgs boson mass 
in the region $(115,720)$ GeV. 
The red, green dots correspond to $\alpha=45^\circ, 20^\circ$, respectively.
The dots are for $(m_1,m_2)(\GeV) =(30,120),(60,120),(90,120),(120,120),(120,320),
(120,520),(120,720)$ from above for each color.
The SM always predicts a negative $\De T$ for the Higgs mass larger than $m_h = 120$ GeV. 
However, $\De T$ can be either positive or negative in our model. 
The positive $\De T$ can fit to the EWPT data better.

\subsection{Dark matter relic density}

\begin{figure}
\centering
\includegraphics[width=\textwidth]{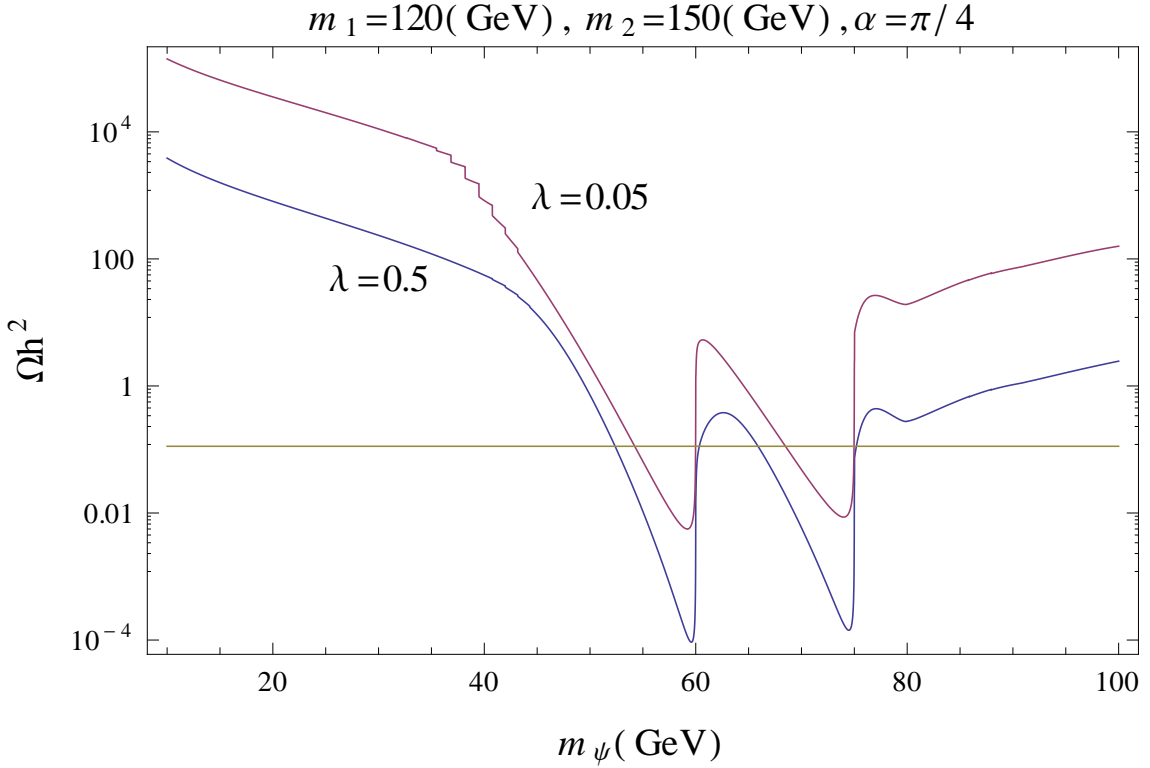}
\caption{Dark matter thermal relic density ($\Omega_{\rm CDM} h^2$) 
as a function of $m_\psi$ for $m_1=120$~GeV, $m_2=150$~GeV, 
$\alpha=\pi/4$ and $\lambda=0.05, 0.5$.}
\label{fig:relic-density}
\end{figure}

The present relic density of cold dark matter, 
$\Omega_{\rm CDM} h^2 \simeq 0.1123 \pm 0.0035$~\cite{Jarosik:2010iu},
is related with the thermally averaged annihilation cross section at freeze-out roughly by 
\beq 
\Omega_{\rm CDM} h^2 \approx \frac{10^{-36} {\rm cm}^2}{\langle \sigma_{\rm ann} v \rangle_{\rm fz}}.
\label{thermal-average-sv}
\eeq
In this paper, we restrict ourselves to the case of $m_\psi < m_i$, for which 
the DM pair  annihilation into $b \overline{b}$ in the $s$-channel becomes dominant.
We can approximate the annihilation cross section 
\bea
\langle \sigma_{\rm ann} v \rangle_{\rm fz}
\approx 10^{-43} \, \l(\la \sin\al \cos\al \over 0.5\r)^2
 \l(m_\psi \over m_1/3\r)^2 
 \l(143 \; {\rm GeV} \over m_1\r)^2 
 \l(T \over m_\psi/25 \r)^2 \;\; {\rm cm^2}.
\eea
The typical value is many orders of magnitude smaller than needed in 
(\ref{thermal-average-sv}).
However, a huge enhancement is possible near the resonance region,
$m_{i} \approx 2 m_\psi$ as can be seen in Fig.~\ref{fig:relic-density}.
The figure shows the present dark matter relic density as a function of 
$m_\psi$ for $m_1 =120 \GeV$, $m_2=150 \GeV$, $\alpha=\pi/4$ 
(maximal mixing) and  $\lambda=0.05, 0.5$.
In the figure, the double-dip is due to two $s$-channel resonances
near $m_\psi = m_i/2$ ($i=1,2$).  We adapted the micrOMEGAs package~
\cite{Belanger:2008sj} to our model for numerical calculation.

\subsection{Direct detection}

For the CDM in the mass range
\beq
m_\psi = \mathcal{O}(10-100) \GeV ,
\eeq
there is a strong upper bound on the spin-independent (SI) dark matter-proton 
scattering cross section from various direct detection experiments
~\cite{Aprile:2011hi}:
\beq 
\sigma_{\rm SI} \lesssim 10^{-44} {\rm cm}^2.
\eeq
The spin-independent (SI) elastic scattering cross section for a Dirac fermion 
dark matter to scatter off a proton target is given by
\beq
\sigma_p \approx \frac{m_r^2}{\pi} \lambda_p^2
\eeq
where $m_r$ is the reduced mass $m_r = m_\psi m_p/(m_{\psi} + m_p)$, 
and $\lambda_p$ is given by 
\beq
\frac{\lambda_p}{m_p} = \sum_{q=u,d,s} f^{(p)}_{Tq} \frac{\la_q}{m_q} + 
\frac{2}{27} f^{(p)}_{Tg} \sum_{q=c,b,t}\frac{\la_q}{m_q} . 
\eeq
The couplings $\lambda_q$'s describe the effective SI four fermion interactions  
of the quarks and the dark matter, and are  given by 
\beq
\frac{\la_q}{m_q} = \frac{\lambda \sin\al \cos\al}{v_H} 
\left( \frac{1}{m_{1}^2}-\frac{1}{m_{2}^2} \right) .
\label{eq:DM_quark_coupling}
\eeq
The parameter $f$'s are defined by the following matrix elements 
\[
m_p f^{(p)}_{Tq} \equiv \langle p|m_q \bar{q}q|p \rangle
\] 
for $q=u,d,s$, and $f^{(p)}_{Tg} = 1-\sum_{q=u,d,s} f^{(p)}_{Tq}$. 
The numerical values of the hadronic matrix elements $f^{(p)}_{Tq}$ we used
are~\cite{Belanger:2008sj}
\beq
f^{(p)}_{Tu}=0.023, \ \ f^{(p)}_{Td}=0.033, \ \
f^{(p)}_{Ts} = 0.26.
\eeq
Note that recent study of these parameters in the lattice QCD yields 
somewhat lower values \cite{Bali:2011rs}. In case we adopt these new numbers, 
the constraints from the direct detection experiments will become milder. 

After all,  for the case $m_\psi \gg m_p$, we find
\bea \label{d-sigma-th}
\sigma_p 
&\simeq& 5 \times 10^{-9} \ {\rm pb} \left( \frac{143 \GeV}{m_1} \right)^4 
\left( 1 - \frac{m_1^2}{m_2^2} \right)^2 \left( \frac{\lambda \sin \theta 
\cos \theta}{0.1} \right)^2 .
\eea

\begin{figure}
\centering
\includegraphics[width=0.7\textwidth]{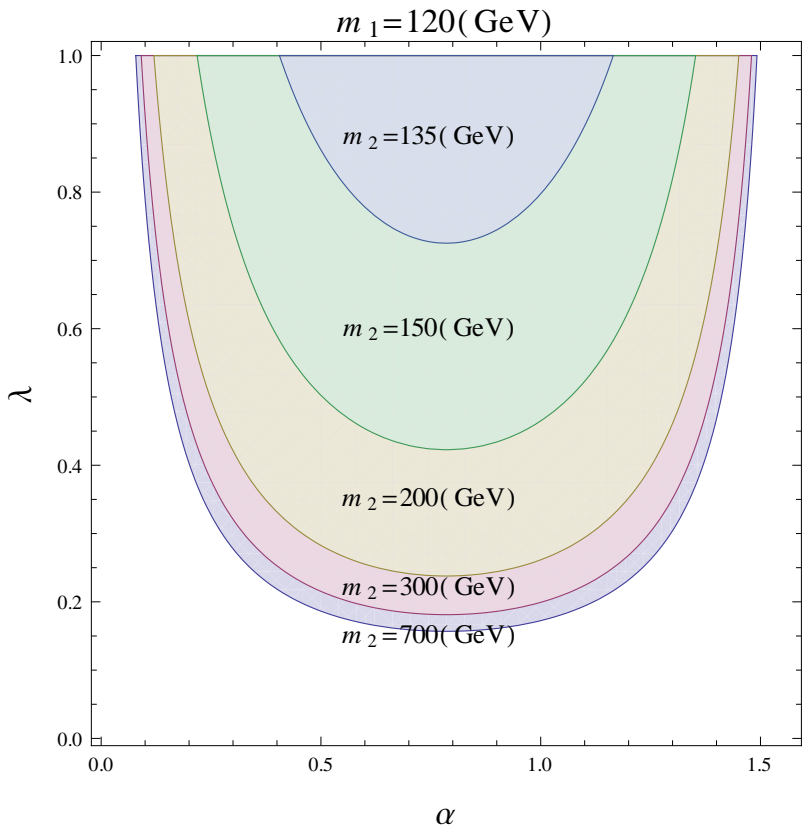}
\caption{Region in the $(\al,\la)$ plane that is excluded by $\si_p < 10^{-8}$ (pb). 
For this plot, we fixed $m_\psi = 70 \GeV$, $m_1 = 120 \GeV$. 
The $m_2$ values used are shown in the plot.  As $m_2$ becomes closer to $m_1$, 
the $\si_p$ constraint gets weaker, as explained in the text.
}
\label{fig:sigmaP_al_la}
\end{figure}

A remark is in order here. 
One might think that the strong constraint 
on the SI cross section from DM direct detection experiments exclude a 
possibility of the SM Higgs decay mode into a pair of DM, since the DM 
couplings to the Higgs boson is constrained by the direct detection experiments.
However this is not true in general. 
For the case of vector [17] and scalar [6] Higgs portal DM, the invisible 
branching ratios can be as large as 80\% and 60\%, respectively, 
even without the singlet scalar messenger $S$.  
In our model, there are two Higgs-like scalar bosons, mixtures of $h$ and $s$. 
Then, as can be seen in (\ref{eq:DM_quark_coupling}), there would be 
a destructive interference  between $H_1$ and $H_2$ contributions to 
the scattering amplitude due to orthogonality of the Higgs mixing matrix, 
which is a very generic aspect in case there are extra singlet scalar bosons that 
can mix with the SM Higgs boson.
Hence, for regions $m_2 - m_1 \ll m_1$, a cancellation occurs in $\si_p$ 
and even the large $\la$, $\al$ regions are only weakly constrained.  
This feature is shown in Fig.~\ref{fig:sigmaP_al_la}, where we show a region in 
the $(\al,\la)$ plane that is excluded by the upper bound of 
$\si_p= 10^{-8}$ (pb).  We also note that  $\sigma_p$ and 
$\langle \sigma_{\rm ann} v \rangle_{\rm fz}$ are not strongly correlated
near the Higgs resonance where the relic density can be explained.
This helps to evade the strong  bound on $\sigma_p$ while accommodating
the correct CDM density in the universe.   
As we'll see below,  this opens up a very interesting parameter space for 
Higgs boson search at the LHC,  making one or two of the Higgs-like scalar 
bosons can decay into a pair of DM's with a substantial invisible branching ratio(s).   

\section{Implications for the Higgs search at the LHC}

In this section, we investigate the allowed parameter space, 
taking into account of all the constraints discussed in the previous section,  
and see if it is possible to discover the  Higgs(es) at the LHC.
We study the following three benchmark scenarios classified 
according to the Higgs mass relations:
\begin{itemize}
\item Scenario 1 (S1): $m_1 \sim 120$~GeV $\ll m_2$ 
\item Scenario 2 (S2): $m_1 \sim m_2 \sim 120$~GeV
\item Scenario 3 (S3): $m_1 \ll m_2 \sim 120$~GeV
\end{itemize}
\subsection{S1: $m_1 \sim 120$~GeV $\ll m_2$}
\begin{figure}
\centering
\includegraphics[width=0.8\textwidth]{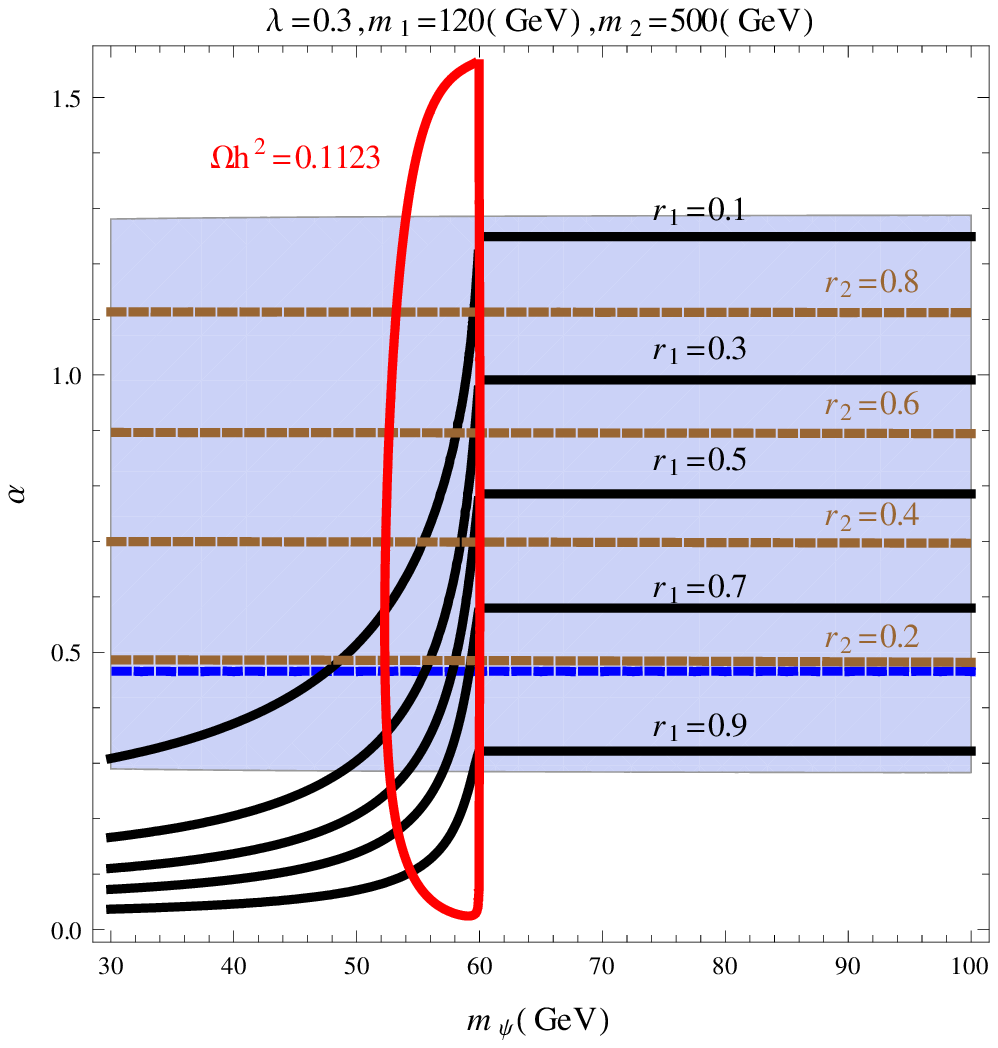}
\caption{Contour plot in $(m_\psi,\alpha)$ plane. We fixed $\la=0.3$, 
$m_1 = 120$~GeV, and
$m_2 = 500$~GeV. The red line represents $\Om_{\psi} h^2 = 0.1123$. 
The sky blue region is excluded by $\si_p <1 \times 10^{-8}$ (pb) 
obtained by XENON100. The region above the dashed blue line is ruled out
by EWPT at 95\% CL.
The solid (dashed) black (brown) lines show reduction factors for 
$H_{1(2)}$:
0.1, 0.3, 0.5, 0.7, 0.9 (0.2, 0.4, 0.6, 0.8) from above (below). 
}
\label{fig:mpsi_alpha}
\end{figure}
In this case, it turns out that the DM direct detection constraint is very strong 
as we can expect from Fig.~\ref{fig:sigmaP_al_la}, and only small $\lambda$ 
is allowed for the mixing angle $\alpha \sim O(1)$.
This can also be seen in Fig.~\ref{fig:mpsi_alpha} where we show a contour 
plot in $(m_\psi,\alpha)$ plane.
We fixed $\la=0.3$, $m_1 = 120$ GeV, and $m_2 = 500 \GeV$ for this plot.
The blue region is excluded by the non-observation of DM at direct detection 
experiments.  The region bounded by the red curve satisfies 
$\Om_\psi h^2 \leq 0.1123$, and the equality holds on the boundary.
The solid black (dashed brown) lines show reduction factors for $H_1$ ($H_2$): 
0.1, 0.3, 0.5, 0.7, 0.9 (0.8, 0.6, 0.4, 0.2) from above.
Since the heavier Higgs is quite heavy with $m_2 = 500$ GeV, 
the contributions to $S,T,U$ parameters  become large for large $\al$ 
as can be seen in   Figs.~\ref{fig:STU_m2_al} and \ref{fig:STU}. 
The region above the dashed blue line is ruled out by EWPT at 95\% CL.
We can see that invisible Higgs decay BR can be sizable, 
reducing signal strength significantly for $m_\psi <60$~GeV.
Considering all these constraints, we find only the lower blank region is
allowed and we get $r_1 \gtrsim 0.5$, $r_2 \lesssim 0.1$.
As a consequence, the possibility of detecting $H_2$  at the LHC would be 
closed unless one can achieve a substantial increase in the integrated luminosity. 
However, if the Higgs cascade decay channel $H_2 \to H_1 H_1$ is
kinematically allowed,  it is still possible to detect $H_2$ via exotic Higgs 
decay channels via $H_2 \to H_1 H_1 \to b b \bar{b} \bar{b}, b \bar{b} 
\tau^+ \tau^-,  \tau^+ \tau^+ \tau^- \tau^-$, which clearly deserves 
more detailed study at the LHC.
In this region the allowed DM direct detection cross section is 
just below the current experimental sensitivity and the current or near-future 
direct detection experiments can probe this scenario.

In fact, these features remain the same even if the CDM is very heavy so that
both Higgs bosons cannot decay into a pair of CDM's. For example, one can
achieve thermal relic density with $( m_1 , m_2 ) = (120, 300)$ GeV, 
$m_\psi = 700$ GeV, $\alpha \sim 0.2$ and $\lambda \sim 1.5$.
For such a heavy DM, the constraint from the direct detection is rather weak,
and one can have rather large $\lambda \sim 1.5$.  Such a heavy CDM can
accommodate the PAMELA excess if it decays via higher dimensional operators. 
Also, the vacuum stability condition should be modified from the SM case,
since there are additional contributions from scalar boson $s$ and the 
fermionic DM $\psi$. These two contributions will compete and can change 
the stability bound curve. It would remain to be seen if the Higgs boson with 
mass around 120 GeV might be still consistent with no new physics up to Planck 
scale or not.  
These issues are somewhat outside the main subjects of this work, and 
will be addressed in more detail in a separate publication \cite{work}.

\subsection{S2: $m_1 \sim m_2$}
\begin{figure}
\centering
\includegraphics[width=0.8\textwidth]{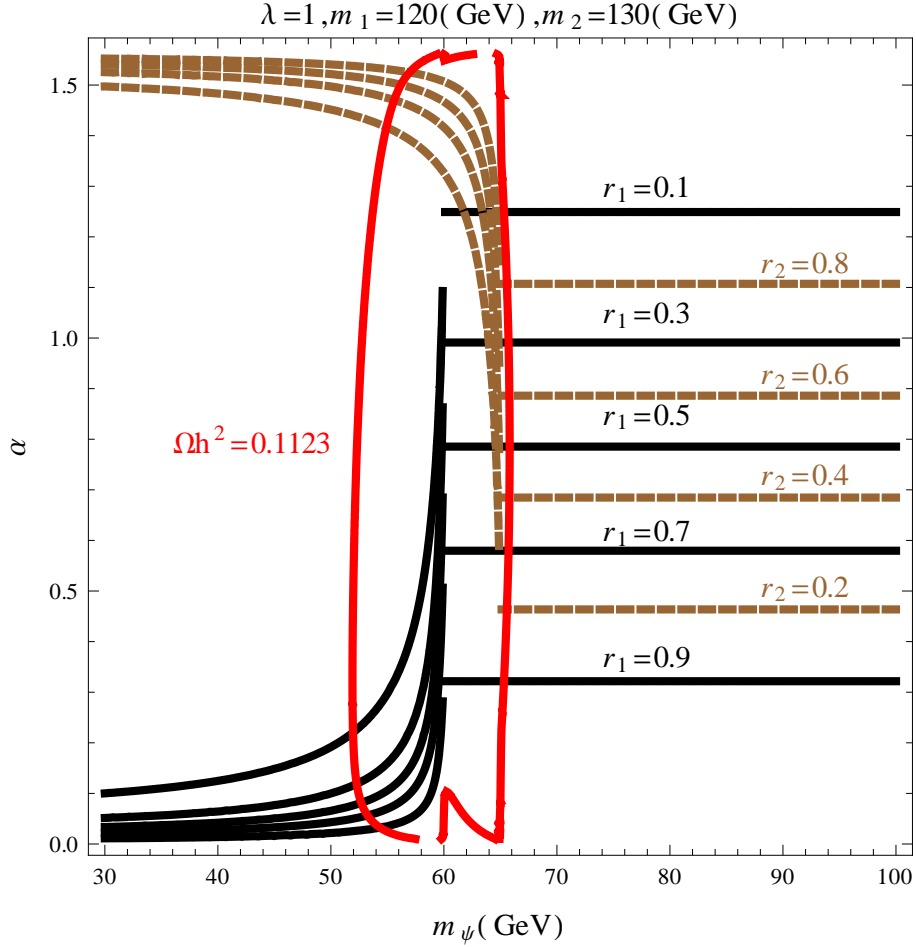}
\caption{Contour plot in $(m_\psi,\alpha)$ plane. We fixed $\la=1$, 
$m_1 = 120$~GeV, $m_2 = 130$~GeV. 
Others are the same with Fig.~\ref{fig:mpsi_alpha}.
Note that the direct detection bound and the EWPT do not constrain 
this scenario for this choice of parameters.
}
\label{fig:mpsi_alpha_degenerate}
\end{figure}
\begin{figure}
\centering
\includegraphics[width=0.8\textwidth]{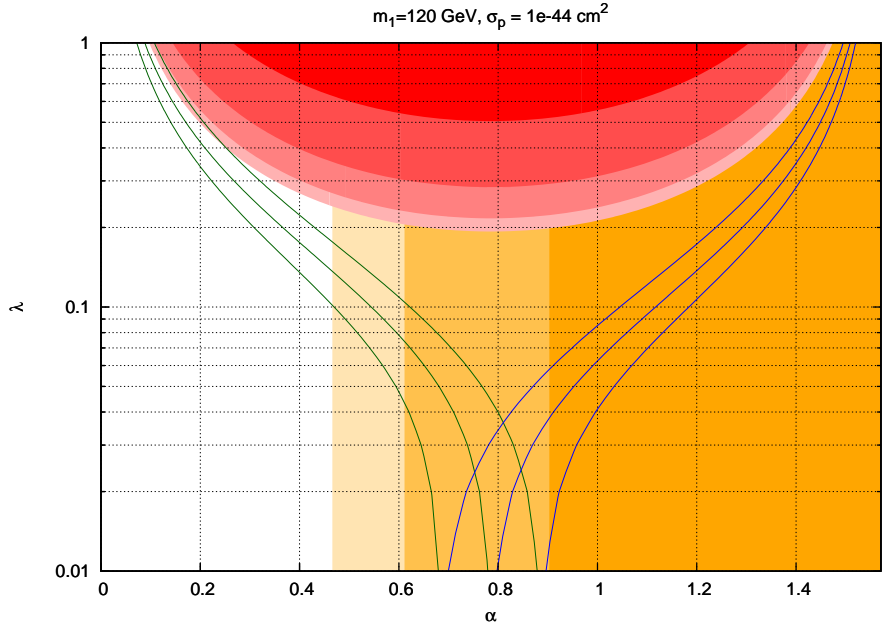}
\caption{Contour plot in $(\alpha,\lambda)$ plane for the case S1 and S2 
with $\sigma_p = 10^{-8}$ (pb) as the upper-bound from direct detection experiments. 
The orange shaded regions are excluded by the constraint from EWPT.
The borders correspond to $m_2=200,300,500 \GeV$ from right.
The reddish shaded regions are excluded by the upper-bound of dark matter-nucleon 
cross section.
The borders correspond to $m_2=150,200,300,500 \GeV$ from above.
The green lines are $r_1=0.4,0.5,0.6$ from above for $m_\psi=50 \GeV$.
The blue lines are $r_2=0.4,0.5,0.6$ for $m_2=150 \GeV$ from above.
}
\label{fig:lambda_alpha_largem1}
\end{figure}

In the S2, the direct detection cross section is drastically suppressed due 
to the destructive interference between $H_1$ and $H_2$. 
The suppression factor is proportional to $(m_2-m_1)^2/m_1^2$.
As we noticed above the cancellation does not necessarily mean too small
annihilation cross section for the relic density, or a small branching ratio 
for invisible Higgs decays.
This is especially true near the $s$-channel resonance region.
This feature can be clearly seen in Fig.~\ref{fig:mpsi_alpha_degenerate}
where there is no constraint from the DM direct detection results (namely, 
no blue box compared with Fig.~5)  and the relic density can be still explained.
For this plot we fixed $\la=1$ which is quite large and makes the
invisible decays quite effective once kinematically allowed.
For the Higgs boson masses,  we chose $m_1 = 120$~GeV, $m_2 = 130$~GeV.
For $m_\psi > 65 \GeV$ Higgs invisible decays are closed and 
a simple sum rule holds between the two reduction factors 
$r_{1,2}$ as can be seen from (\ref{eq:rf}) 
(see the discussion in Sec.~\ref{sec:LHC} and Fig.~\ref{fig:r1r2}):
\bea
  r_1 + r_2 = 1.
\eea
This is because the Higgs invisible and splitting decay modes are kinematically 
closed and the event reduction occurs only at the Higgs boson productions.
If $m_\psi \gtrsim 65 \GeV$ and $\al$ is nearly maximal, both $r_1$ and $r_2$ 
are close to $0.5$. So both Higgs bosons can be observed at the LHC by the standard
Higgs search method. This would be a clear signal for S2 as 
well as the existence of additional singlet scalar.
 
Again from Fig. \ref{fig:lambda_alpha_largem1}, it is clear that if $m_2$ becomes 
close to $m_1$, wide ranges of $\lambda$ and $\alpha$ are allowed.
If $m_2= 150 \GeV$, for example, the constraint from direct detection becomes 
almost irrelevant due to the destructive interference between $H_i$'s contributions 
(see \eq{eq:rf}). The constraint from the EWPT is also irrelevant.
In this case, both of $r_1$ and $r_2$ can be close to 0.5 if the mixing is nearly maximal, 
but it happens only for $\lambda \ll 1$ implying that invisible decay branching ratios for 
the two Higgs bosons are negligible. 
This would be another case where we can see  both Higgses.

\subsection{S3: $m_1 \ll m_2 \sim 120$~GeV}
\begin{figure}
\centering
\includegraphics[width=0.8\textwidth]{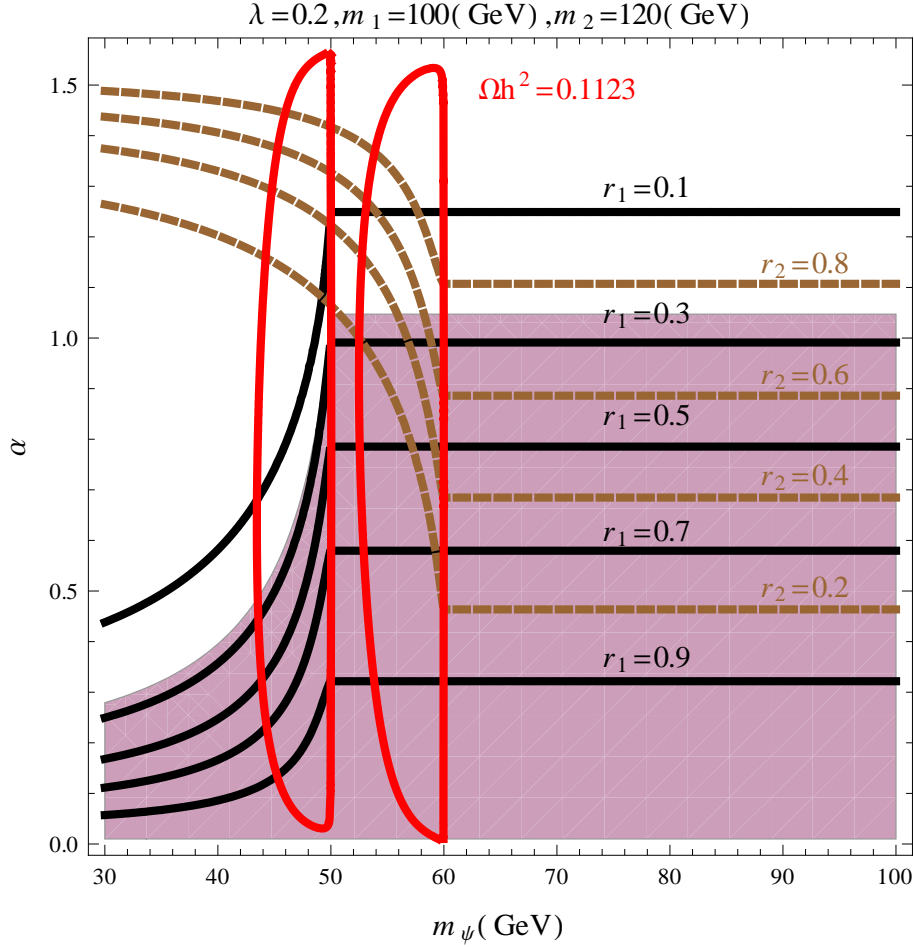}
\caption{Contour plot in $(m_\psi,\alpha)$ plane for the case S3 with 
$\lambda=0.2$, $m_1 = 100$~GeV, $m_2 = 120$~GeV.
The purple region is excluded by the LEP Higgs search bound
$m_H^{\rm SM} > 114.5$~GeV.
Others are the same with Fig.~\ref{fig:mpsi_alpha}.
Note that the direct detection bound and the EWPT do not constrain 
this scenario for this choice of parameters.
}
\label{fig:mpsi_alpha_inv}
\end{figure}
\begin{figure}
\centering
\includegraphics[width=0.8\textwidth]{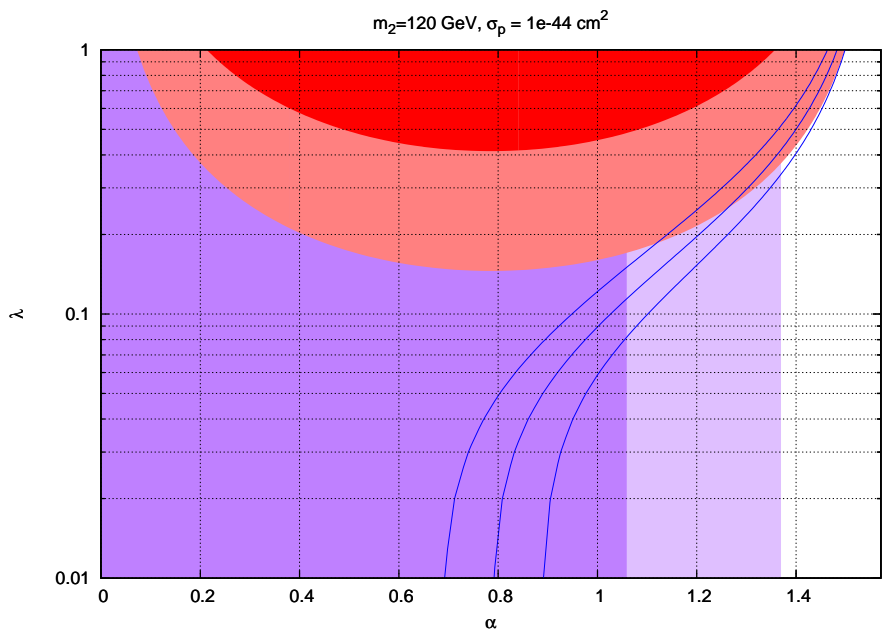}
\caption{Contour plot in $(\alpha,\lambda)$ plane for the case S3 with 
$\sigma_p = 10^{-8}$ (pb) as the upper-bound from direct detection 
experiments.  The borders of color-shaded regions correspond to 
$m_1=80,100 \GeV$ from right (below) for purple (reddish) regions.
The purple shaded regions excluded by the constraint on $r_1$ from LEP data.
The borders correspond to $r_1=0.04,0.24$ from right.
The reddish shaded regions are excluded by the upper-bound of dark 
matter-nucleon cross section.
The blue lines are $r_2=0.4,0.5,0.6$ from left for $m_\psi=50 \GeV$.}
\label{fig:lambda_alpha_inv}
\end{figure}
In this case, for $m_1 \gtrsim 50 \GeV$, the constraints from 
$S$, $T$ and $U$ oblique parameters of EWPT become irrelevant 
as shown in the right panel of Fig.~\ref{fig:STU_m2_al}. 
The most stringent constraint comes from the LEP Higgs search bound.
In Fig.~\ref{fig:mpsi_alpha_inv}, we show the case of $\lambda=0.2$,  
$m_1=100$~GeV, and $m_2 =120$~GeV with  the LEP
excluded region shown in purple which corresponds to $r_1 \ge 0.25$.
The current constraint from the DM direct detection experiments is satisfied 
for the parameter set chosen for the plot.
We see that $r_2$ can take any value from 0 to 1.
If $m_2 < 2 m_\psi$ and the invisible decay of $H_2$ is closed, 
only $r_2 \gtrsim 0.8$ is allowed. That is, if $H_2$ is observed with
$r_2 <0.8$ we can conclude that the reduction of the signal strength
is due to the Higgs decay to the hidden sector dark matter.
In Fig. \ref{fig:lambda_alpha_inv}, we can find the dependence of $r_2$ 
on $\lambda$ due to the constraint from direct 
detection. 
For example, if $m_1 \lesssim 80 \GeV$ only $r_2 \gtrsim 0.6$ is allowed.

\subsection{The LHC reach}
\label{sec:LHC}
\begin{figure}
\centering
\includegraphics[width=0.7\textwidth]{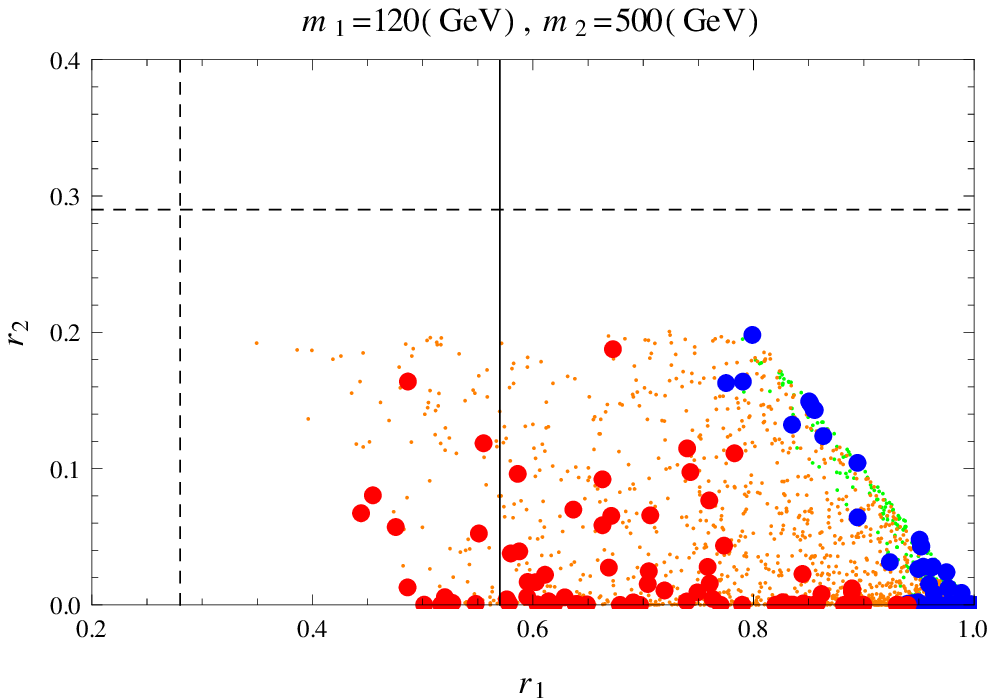}
\includegraphics[width=0.7\textwidth]{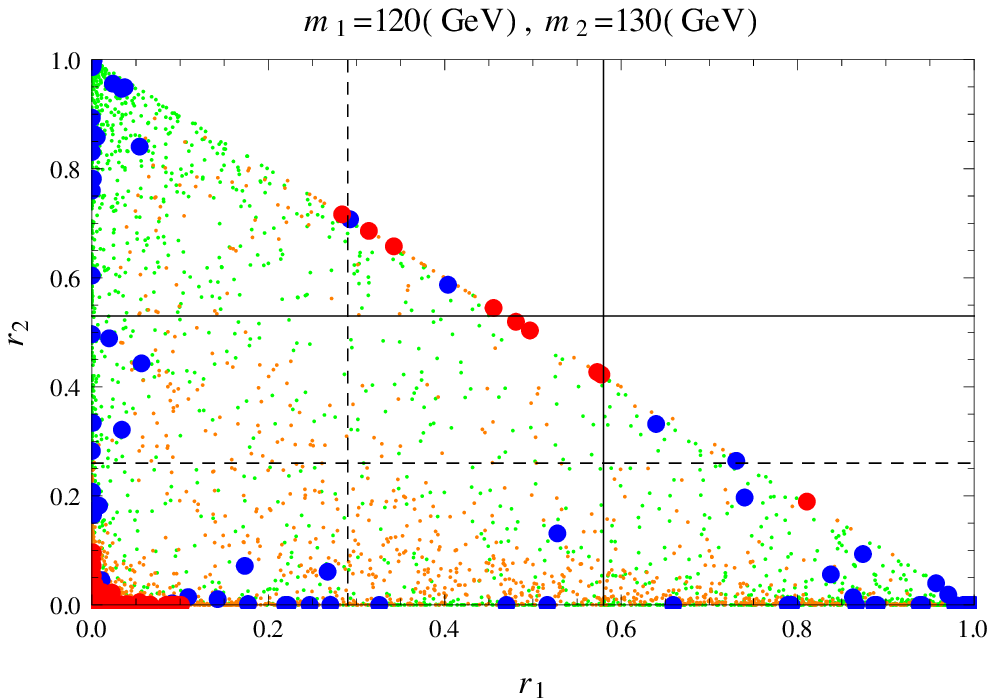}
\includegraphics[width=0.7\textwidth]{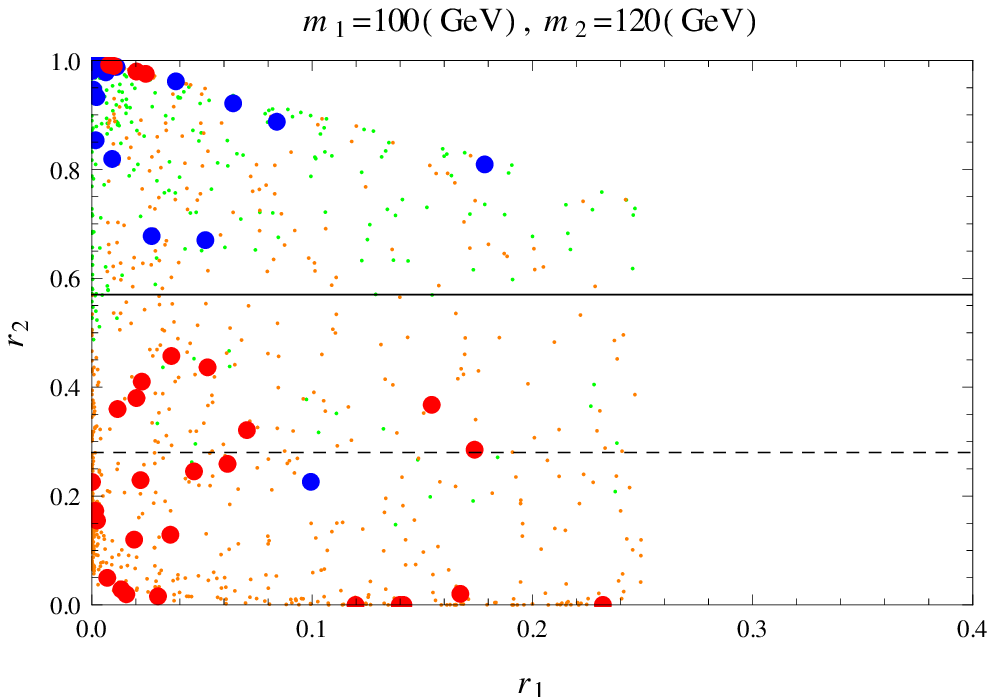}
\caption{Scatter plot in $(r_1,r_2)$ plane for the scenario 
S1, S2 and S3 (from above). The region that the LHC can probe 
at 3$\sigma$ level  with 5 (10) fb$^{-1}$ 
luminosity is represented by solid (dashed) line. 
The points represent 4 different cases:
$(\Om_{\rm CDM} h^2)^{3 \si}$, $\si_p^>$ (big red),
$(\Om_{\rm CDM} h^2)^{3 \si}$, $\si_p^<$ (big blue),
$(\Om_{\rm CDM} h^2)^{<}$, $\si_p^>$ (small orange),
and $(\Om_{\rm CDM} h^2)^{<}$, $\si_p^<$ (small green). 
(See the text for more detail).
}
\label{fig:r1r2}
\end{figure}

In this subsection we concentrate on the parameter space that can be 
probed at the LHC in near future.  For this purpose we show scattered 
plots in $(r_1, r_2)$ plane in Fig.~\ref{fig:r1r2}.
We take the Higgs masses as benchmark points for each scenario as follows:
\begin{itemize}
\item S1: $m_1 = 120$ GeV,  $m_2 = 500$ GeV,
\item S2: $m_1 = 120$ GeV,  $m_2 = 130$ GeV,
\item S3: $m_1 = 100$ GeV,  $m_2 = 120$ GeV.
\end{itemize}
We scanned the remaining parameters in the range
\bea
  0 < &\la& < 1, \nl
 10 < &M_\psi& <100 \ \ {\rm GeV}, \nl
 0 < &\al& < \pi/2.
\eea
All the points in the plots satisfy
\begin{itemize}
\item the unitarity condition~(\ref{eq:unitarity}), 
\item the LEP Higgs mass bound~\cite{Barate:2003sz},
\item the EWPT fits at 95\% CL~\cite{Barger:2007,Baak:2011ze},
\item the direct search bound of DM by XENON100, $\sigma_p < 10^{-8}$ pb,
\item the relic density of DM, $\Om_{\rm CDM} h^2 < 0.1228$.
\end{itemize}
We can divide the $\si_p$ into two regions:
\bea
 \si_p^>: 10^{-9} <  \sigma_p < 10^{-8}, \quad
 \si_p^<: \sigma_p < 10^{-9}, 
\eea
where the former region can be probed in near future direct search
experiments. The relic density is also divided into two regions:
\bea
 (\Om_{\rm CDM} h^2)^{3 \si}: 0.1018 < \Om_{\rm CDM} h^2 < 0.1228, \quad
 (\Om_{\rm CDM} h^2)^{<}:  \Om_{\rm CDM} h^2 < 0.1018.
\eea
where the former is the WMAP 3$\sigma$ allowed region.
The different colors and sizes of the points represent different 
regions of $\si_p$ and $\Om_{\rm CDM} h^2$ as table \ref{tb:color-scheme}.
\begin{table}[h] 
\begin{center}
\begin{tabular}{c|cc}
\hline
  & $\si_p^>$ & $\si_p^<$ \\
\hline
$(\Om_{\rm CDM} h^2)^{3 \si}$ & big red &  small orange\\
$(\Om_{\rm CDM} h^2)^{<}$ & big blue & small green \\
\hline
\end{tabular}
\end{center}
\caption{Color Schemes for $\sigma_p$ and $\Omega_{\rm CDM} h^2$}
\label{tb:color-scheme}
\end{table}

The region that the LHC can probe at 3$\sigma$ level  with 5 (10) fb$^{-1}$ 
luminosity is represented by solid (dashed) line~\cite{CMS-TDR,ATLAS-TDR}.
The S1 scenario can be tested fully at the LHC with 10 fb$^{-1}$
by observing $H_1$. 
In the case of S2 the LHC may see both Higgs bosons with the standard 
search strategy.
However, there are still some points which the LHC has difficulty to 
find two Higgs bosons.  These are the points near the origin 
($r_1 \approx r_2 \approx 0$)  where the invisible decays becomes dominant.
In S3 the region with small $r_2 (<0.24)$ can not be probed with the standard
decay channels. However, once $H_2 \to H_1 H_1$ is open, this region can also be
tested at the LHC.

\subsection{The implications of the recent LHC reports on our model}
Recently the ATLAS~\cite{ATLAS_Dec13} and the CMS~\cite{CMS_Dec13} 
at the LHC reported that the allowed SM Higgs mass  range is further 
constrained to be $115 < m_h < 131$~GeV (ATLAS)  and $m_h < 127$~GeV (CMS) 
and they also saw a hint with the mass in the range $124 < m_h < 126$~GeV.
The analysis in our paper with a Higgs mass 120~GeV does not change much 
by changing it to 125~GeV.
In our model the Higgs with mass $~125$~GeV can be either $H_1$ or $H_2$.
Assuming $r \gtrsim 0.6$ to be discovered with 3$\sigma$ significance, we can
see that the possible Higgs signal at the LHC should be $H_1$ ($H_2$) for the
scenario S1 (S3). In these models the remaining Higgs lies beyond the reach of the LHC.
For the scenario S2, the LHC may have seen either $H_1$ or $H_2$. 
The other Higgs may or may not be probed at the LHC with more luminosity.
In summary, if we consider the recent results for the  SM Higgs boson 
at the LHC as a real singal for the Higgs bosons around 125 GeV Higgs boson 
with $r_i \sim 1$  in our model, 
the other Higgs boson can be either light  or heavy, and 
it would be very difficult to discover it at the LHC 
\footnote{Similar discussions were presented in Refs.~
\cite{Kadastik:2011aa,Djouadi:2011aa}  after the 
ATLAS and the CMS reported the new results on the SM Higgs boson.}.

\section{Conclusions}
In this paper we considered a simple extension of the SM where the fermionic 
DM in the hidden sector can interact with the SM sector through Higgs portal.
A new singlet scalar has been introduced as a messenger between the hidden sector 
and the SM sector. It mixes with the SM Higgs boson and behaves like a second 
Higgs boson.   This opens up a new possibility that the Higgs search program 
can be quite  non-standard.

Considering the constraints mainly from the LEP Higgs search bound,
electroweak precision observables, thermal dark matter density and 
dark matter direct detection experiments, we investigated the possible 
Higgs search scenarios in three categories:
\begin{itemize}
\item Scenario 1 (S1): The light Higgs boson $H_1$ has mass 
$\sim 120 \GeV$ and the heavier one $H_2$ is much heavier than $H_1$.
\item Scenario 2 (S2): The two Higgs bosons are almost degenerate in mass: 
$ m_2 - m_1 \lesssim 20 \GeV$. We assume $m_1 = 120 \GeV$.
\item Scenario 3 (S3): The heavy Higgs boson $H_2$ has mass $\sim 120 \GeV$.
\end{itemize}

For S1, the constraints from EWPT and DM direct search data severely restrict 
the discovery likelihood of the heavier Higgs.
The mixing angle in this case is pushed to small value causing significant 
signal reduction for the $H_2$ discovery (see the top plot in Fig.~10).   
However, $r_1 \gtrsim 0.4$ is possible while $\lambda$ is still not so small.
Hence $H_1$ can be discovered even if it has a sizable invisible branching ratio.
Most of the points in the top plot of Fig.~10 are also sensitive to near-future 
DM direct search experiements.
Especially, the lighter Higgs may behave like the SM Higgs boson if 
$r_1 \sim 1$,  and then can be discovered at the LHC.
On the other hand, the heavier Higgs is very difficult to observe at the LHC,
since $r_2 \sim 0$ for $r_1 \sim 1$.
It would be important to measure $r_1$ as precisely as possible, and see 
if it deviates from the SM value or not in order to test our model. 
However, because of theoretical and experimental uncertainties, 
it would be a difficult  job, especially for $m_1 \sim 120$ GeV.
If one could improve the sensitivity on $r_2$ from $H_2 \rightarrow WW, ZZ$ 
below $r_2 < 0.1$, one might be able to find out the heavier Higgs boson 
at the LHC with higher luminosity. 

On the other hand, for S2, the constraints from the EWPT become weak.
Interestingly the constraint from direct detection experiments is 
drastically alleviated due to a strong cancellation in the $\sigma_p$.
As the result, there is an ample room to discover one or both of the Higgs 
particles with the standard Higgs search method at the LHC, if the invisible decays of 
$H_i$ is kinematically closed and the relation $r_1 + r_2 = 1$ holds.
There is also possibility that both Higgs bosons may not be seen at the LHC,  
if invisible decay modes become significant and $r_1 , r_2 \sim 0$ 
(see the middle plot in Fig.~10).  In this particular case, both Higgs bosons
would not be discovered at the LHC, and also no new particles appear 
below $\sim 1$ TeV. This would seem to be in conflict with the theorem 
by Lee, Quigg and Thacker \cite{bwlee}. However in our model, it is possible 
that we observe no particles including Higgs bosons and other new resonances 
at the electroweak scale, without violating the unitarity of the longitudinal 
weak gauge boson scattering amplitudes. 

In the case S3, the most stringent constraint is from LEP Higgs mass bound for 
the lighter Higgs. As a consequence, it is difficult to detect $H_1$
while $H_2$ can be seen at the LHC (see the bottom plot in Fig.~10). 
The DM direct detection experiment may also see a signal.

In summary, in the fermionic DM model with Higgs portal,
it is easy to explain the DM relic density while satisfying
the DM direct detection bounds from the cancellation.
Two Higgs-like bosons can decay into dark matter pair, thereby the signals 
for the Higgs bosons being reduced significantly.  
We have a big chance of discovering one or both of Higgs particles 
at the LHC in the near future if two Higgs bosons have hierarchical mass spectra.  
On the other hand, we may find no Higgs  bosons or new resonances 
if $m_1 \sim m_2$ and $r_1 , r_2 \sim 0$.
Even if we find no Higgs boson at the LHC, it would not imply that perturbative
unitarity in the longitudinal weak gauge bosons is violated, since our model
is renormalizable and respects unitarity, still having a possibility of no visible 
effects at the LHC at all.
It would remain to be seen which route is realized in the nature.

\section*{Acknowlegements}

We are greatful to Suyong Choi, Dong Hee Kim and Soo Bong Kim 
for useful discussions. 
This work of PK was supported in part by the National Research Foundation (NRF) 
through Korea Neutrino Research Center (KNRC) at Seoul National University.

\appendix
\section{The loop functions for the $S, T$ and $U$ parameters}
\label{sec:appendix}
The loop functions for the $S, T$ and $U$ parameters are listed below:
\bea
f_T(x) &=& \frac{x \log x}{x-1}, \nl
f_S(x) &=& \left\{
\begin{aligned}
& \frac{1}{12} \left[-2 x^2+9 x
+\left(x^2-6 x-\frac{18}{x-1}+18\right) x \log x \right.\\
& \left. +2 \sqrt{(x-4) x} \left(x^2-4 x+12\right)
  \left(\tanh
   ^{-1}{\frac{\sqrt{x}}{\sqrt{x-4}}}-\tanh
   ^{-1}{\frac{x-2}{\sqrt{(x-4) x}}}\right)\right] (\text{for}\;\; 0 <x <4), \nl
& \frac{1}{12} \left[-2 x^2+9 x+\left(x^2-6 x-\frac{18}{x-1}+18\right) x \log
   x  \right.\\
& \left.+\sqrt{(x-4) x} \left(x^2-4 x+12\right) \log \frac{1}{2}
   \left(x-\sqrt{(x-4) x}-2\right)\right] (\text{for}\;\; x >4).
\end{aligned}\right.
\eea

\end{document}